\documentclass[11pt,a4paper,oneside]{article}
\usepackage[T1]{fontenc}
\usepackage[ansinew]{inputenc}
\usepackage[english]{babel}
\usepackage{amsfonts}
\usepackage{amsmath}
\usepackage{bm}
\usepackage{array}
\usepackage{amsthm}
\usepackage{amssymb}
\usepackage{graphicx}
\usepackage{braket}
\usepackage{verbatim}
\usepackage[table]{xcolor}
\usepackage{caption}
\usepackage{cite}
\usepackage{textcomp}
\usepackage{url}
\usepackage{hyperref}
\hypersetup{
    colorlinks=true,
    linkcolor=black,
    citecolor=black
    }
\usepackage{multicol}
\newcommand{\be}{\begin{equation}}
\newcommand{\ee}{\end{equation}}
\definecolor{pinegreen}{rgb}{0.0, 0.47, 0.44}

\renewcommand{\d}{\mbox{${\rm d}$}}

\newtheorem{theorem}{Theorem}
\newtheorem{corollary}{Corollary}

\theoremstyle{definition}

\theoremstyle{remark}
\newtheorem{remark}{Remark}

\newcommand{\nn}{\nonumber}

\newcommand{\K}{\mathcal{K}}
\newcommand{\T}{\mathcal{T}}

\title{\bf Alternative formulations of the\\ thermodynamics of scalar-tensor theories} %
\author{
Luca~Gallerani$^{1,2}$,
Marcello~Miranda$^{4,5}$
Andrea~Giusti$^{1,6}$
, and Andrea~Mentrelli$^{1,2,3}$
\\
\\
$^1${\em Alma Mater Research Center on Applied Mathematics (AM$^2$)}
		\\
{\em Via Saragozza 8, 40123 Bologna, Italy}
\\
\\
$^2${\em Department of Mathematics, University of Bologna}
	\\
	{\em Piazza di Porta San Donato 5, 40126 Bologna, Italy} 
\\
\\
$^3${\em Istituto Nazionale di Fisica Nucleare (I.N.F.N.), Sezione di Bologna, I.S. FLAG}
	\\
	{\em Viale Berti Pichat 6/2, 40127 Bologna, Italy}
 \\
 \\
$^4$ {\em Scuola Superiore Meridionale,} \\
{\em Largo San Marcellino 10, 
I-80138, Napoli, Italy}
\\
\\
$^5$ {\em Istituto Nazionale di Fisica Nucleare (I.N.F.N.), Sezione di Napoli,}\\ 
{\em Complesso Universitario Monte Sant'Angelo, Edificio G,}\\ 
{\em Via Cinthia, I-80126, Napoli, Italy}
\\
\\
$^6$ {\em Department of Physics and Astronomy, University of Sussex}\\ 
{\em Brighton, BN1 9QH, United Kingdom}
}
\begin{document}
\maketitle
\begin{abstract}
\noindent We explore alternative formulations of the analogy between viable Horndeski gravity and Eckart's first-order thermodynamics. We single out a class of identifications for the effective stress-energy tensor of the scalar field fluid that, upon performing the imperfect fluid decomposition, yields constitutive relations that can be mapped onto Eckart's theory. We then investigate how different couplings to Einstein's gravity, at the level of the field equations, can affect the thermodynamic formalism overall. Last, we specialize the discussion to the case of ``traditional'' scalar-tensor theories and identify a specific choice of the coupling function that leads to a significant simplification of the formalism.
\end{abstract}
\newpage
%
%
%
\section{Introduction and Main Results}
\label{sec-1}

A seemingly strict relationship between gravitational theories and thermodynamics has been attracting attention since the 1970s, beginning with the modern formulation of black hole thermodynamics (see, {\em e.g.},~\cite{Hawking:1976de, Bekenstein:1973ur}, and~\cite{Page:2004xp}) and evolving into the {\em thermodynamics of spacetime}~\cite{Jacobson:1995ab, Eling:2006aw}. Jacobson's approach~\cite{Jacobson:1995ab}, in particular, allows for a derivation of the Einstein field equation as an equation of state, while modified theories of gravity emerge from a {\em non-equilibrium formulation of spacetime thermodynamics}~\cite{Eling:2006aw}. These general considerations then inspired an alternative perspective on the relationship between gravity and thermodynamics, known as the {\em thermodynamics of scalar-tensor gravity}~\cite{Faraoni:2021lfc, Faraoni:2021jri} (see also~\cite{Giardino:2023ygc} for a review). This latter approach relies on an effective fluid treatment of the contributions of the additional scalar field degree of freedom in the field equations of scalar-tensor theories, rather than on a thermodynamic description of the spacetime.

For the sake of clarity, let us expand on the generalities of the thermodynamics of scalar-tensor gravity~\cite{Faraoni:2021lfc, Faraoni:2021jri}. Let us consider a generic scalar-tensor theory of gravity with a (total) action
\begin{equation}\label{eq:action}
S\left[g_{ab},\phi, \psi\right]=S^{\rm(g)}\left[g_{ab},\phi\right]+S^{\rm(m)}\left[g_{ab},\psi\right]\,,
\end{equation}
where $g_{ab}$ denotes the metric tensor field, $\phi$ is a scalar field, $S^{\rm(m)}\left[g_{ab},\psi\right]$ denotes the action of a collection of matter fields dubbed by $\psi$ which are minimally coupled to the metric tensor field, and $S^{\rm(g)}\left[g_{ab},\phi\right]$ is the Jordan frame gravitational action of the theory. Taking the functional derivatives of $S^{\rm(g)}\left[g_{ab},\phi\right]$, with respect to $g_{ab}$ and $\phi$, one can define the following quantities
\be \label{fieldeq}
\mathcal{E}_{ab} :=\frac{2}{\sqrt{-g}}\frac{\delta S^{(g)}}{\delta g^{ab}}\,,\quad \mathcal{J}:=\frac{\delta S^{(g)}}{\delta \phi} \, ,
\ee
while the functional derivative of the matter action with respect to $g_{ab}$ yields the energy-momentum tensor of matter, {\em i.e.},
\be 
T^{(m)}_{ab} := -\frac{2}{\sqrt{-g}}\frac{\delta S^{(m)}}{\delta g^{ab}} \, .
\ee
The equations of motion of the system described by the total action $S\left[g_{ab},\phi, \psi\right]$ then read, in general,
\be 
\label{eq-1}
\mathcal{E}_{ab} = T^{(m)}_{ab} \, ,
\ee
\be
\mathcal{J} = 0 \, ,
\ee
\be
\frac{\delta S^{(m)}}{\delta \psi} = 0 \, ,
\ee
where the latter gives the equations of motion of each matter field. 

Let us assume that Eq.~\eqref{eq-1} can be rewritten as an effective Einstein equations, {\em i.e.},
\be
\label{eff-einst}
G_{ab} = G_{\rm eff} (\phi) \, T^{\rm (m)} _{ab} + T^{\rm (eff)} _{ab} \, ,
\ee
where $G_{ab}$ is the Einstein tensor, $G_{\rm eff} (\phi)$ denotes the effective gravitational coupling between matter and Einstein's gravity, while $T^{\rm (eff)} _{ab}$ is an effective stress-energy tensor containing all the remaining contributions of $\phi$.
Then, the first step of the procedure developed in~\cite{Faraoni:2021lfc, Faraoni:2021jri} begins with interpreting $T^{\rm (eff)} _{ab}$ as the stress-energy tensor of an {\em effective fluid}~\cite{Deffayet:2010qz, Faraoni:2018qdr}, that in the case of scalar-tensor gravity is often dubbed as $\phi$-fluid.

In order to implement this {\em effective fluid approach} to modified gravity one first needs to observe that $T^{\rm (eff)} _{ab}$ is, by construction, a rank-2 symmetric tensor field. Hence, given a normalised timelike vector field $u^a$, {\em i.e.} $u^a u_a = - 1$, $T^{\rm (eff)} _{ab}$ always admits an imperfect fluid decomposition~\cite{Faraoni:2023hwu}, {\em i.e.},
\be \label{eq:imperf}
T^{\rm (eff)} _{ab} = \rho u_{a}u_{b}+P h_{ab}+2q_{(a}u_{b)} +\pi_{ab}
\ee
where $h_{ab} = g_{ab} + u_a u_b$, $\rho=T^{\rm (eff)}_{ab}u^{a}u^{b}$ denotes the effective energy density, $P=\tfrac{1}{3}\,T^{\rm (eff)}_{ab}h^{ab}$ is the isotropic pressure, $q_{a}=-T^{\rm (eff)}_{cd}u^{c}h^{d}{}_{a}$ is the heat flux density, and $\pi_{ab}=h_{a}{}^{c} h_{b}{}^{d}T^{\rm (eff)}_{cd}-Ph_{ab}$ denotes the anisotropic stress tensor. Assuming a timelike gradient for the scalar field $\phi$, we identify the $4$-velocity of the $\phi$-fluid with 
\be \label{4vel}
u^a := \epsilon \, \frac{\nabla^a \phi}{\sqrt{2X}} \, ,
\ee
where $X := -\frac{1}{2}\nabla_{a}\phi\nabla^{a}\phi$ and $\epsilon =\pm 1$ is introduced 
to ensure that the velocity is future-oriented~\cite{Giusti:2022tgq}, one can then compute the expansion scalar $\Theta := \nabla_a u^a$ and the shear tensor $\sigma_{ab}:=\nabla_{(b}u_{a)}-\tfrac{1}{3}\Theta\,h_{ab}$, namely the kinematic quantities associated with the $\phi$-fluid. The {\em constitutive equations} of the $\phi$-fluid can then be inferred by comparing the expressions of the kinematic quantities with the components of $T^{\rm (eff)} _{ab}$ in the imperfect fluid form.

Specializing our discussion to the {\em viable subclass of Horndeski gravity}, {\em i.e.},
\begin{equation}\label{eq:viable}
    S^{\rm (g)}=\frac{1}{2}\int \d ^4 x \sqrt{-g} \, \Big[ G_{4}(\phi) \, R +G_{2}(\phi,X)-G_{3}(\phi,X)\Box\phi\Big] \, ,
\end{equation}
where $G_i$ are arbitrary functions of $\phi$ and/or $X$, and $\Box \phi := g^{ab} \nabla_a \nabla_b \phi$, one finds that the constitutive laws for the corresponding $\phi$-fluid can be mapped into those of Eckart's first-order thermodynamics~\cite{Giusti:2021sku} (see also~\cite{Miranda:2022wkz}), {\em i.e.}, 
\begin{align}
P &= \bar{P} + P_\text{vis} \, , \nonumber \\
\label{EckartPvis}
P_\text{vis} &= -\zeta \, \Theta 
\\
\label{Eckartq}
q_a &= -{\cal K} \left( h_{ab} \nabla^b {\cal T} + {\cal T} \dot{u}_a \right) \\
\label{Eckartpi}
\pi_{ab} &= - 2\eta \, \sigma_{ab} \, , 
\end{align} 
where $\bar{P}$ denotes the inviscid pressure, $P_\text{vis}$ is the viscous pressure, $\zeta$ is the bulk viscosity coefficient, $\eta$ is the shear viscosity of the fluid, ${\cal K}$ is the thermal conductivity, and ${\cal T}$ denotes the temperature. This formal analogy, and specifically the generalized Fourier law in Eq.~\eqref{Eckartq}, then allows one to identify a notion of {\em temperature of gravity} that reads (see~\cite{Giusti:2021sku, Miranda:2022wkz}) 
\be
\label{temp-grav}
{\cal K} {\cal T} := \frac{\epsilon\sqrt{2X}\left(G_{4\phi}-XG_{3X}\right)}{G_{4}} \, ,
\ee
where $G_{i\phi}$ and $G_{iX}$ respectively denote the partial derivatives of $G_i$ with respect to $\phi$ and $X$, as well as notions of bulk and shear viscosity for the $\phi$-fluid of viable Horndeski gravity.

Note that, throughout this work, we shall adopt the notation of Ref.~\cite{Wald:1984rg}, in which the metric signature is $(-+++)$. Furthermore, units are used in which the speed
of light and $8 \pi G$ (where $G$ denotes Newton's constant) are unity.

\subsection{Statement of the problem and main results}\label{sec-1.1}
From the discussion presented so far it might appear as if the proposed thermodynamic analogy of viable Horndeski gravity with Eckart's thermodynamics (and corresponding implications) is heavily dependent on the splitting performed in the right-hand side of Eq.~\eqref{eff-einst}. In other words, it can appear as though, given a scalar-tensor theory of gravity, different identifications for the effective stress-energy tensor $T^{\rm (eff)} _{ab}$ for the same theory would lead to different constitutive laws for the corresponding effective fluid. The aim of this work is to address this point and determine to which extent of the constitutive laws of the $\phi$-fluid are mapped onto Eckart's thermodynamics, and to investigate possible alternative formulations of this analogy.

The procedure followed in Eq.~\eqref{eff-einst} aims at completely separating the contributions of the scalar field $\phi$ from the matter content of the theory. In other words, the $\phi$-fluid and ordinary matter are seen as completely independent ``fluids''. However, a key distinction between the two is the fact that, while the matter stress-energy tensor is covariantly conserved due to the diffeomorphism invariance of the action, the same does not hold for the effective $\phi$-fluid, indeed
\begin{equation}\label{nonconserved}
    \nabla^{b}T^{\rm (eff)}_{ab} = \frac{G_{4\phi}}{G_{4}^2}\,
    T^{\rm(m)}_{ab} \,\nabla^{b}\phi \, ,
\end{equation}
which is non-vanishing in general. This is a direct consequence of the non-minimal coupling of $\phi$ to gravity. Furthermore, another important conclusion that can be drawn from Eq.~\eqref{eff-einst} and the fluid interpretation of $T^{\rm (eff)}_{ab}$ is the fact that the $\phi$-fluid has a constant coupling to Einstein's gravity, whereas that of matter is gauged by the scalar field $\phi$ though $1/G_4 (\phi)$. This somewhat contradicts the universality of the gravitational interaction in this effective Einsteinian analogy for scalar-tensor gravity. However, $T^{\rm (eff)} _{ab}$ was postulated of this form and nothing prevents us from arbitrarily defining a new effective energy-momentum tensor
\be
\label{eff-tensor-f}
T^{(f)}_{ab} := [f(\phi,X)]^{-1} \, {T}^{\rm (eff)}_{ab} \, ,
\ee
with $f(\phi,X)$ a non-vanishing continuous function of $\phi$ and $X$, so that the effective Einstein equation can be recast as
\be
\label{eff-einst-f}
G_{ab} = \frac{1}{G_{4}} \, T_{ab}^\text{(m)} + f(\phi,X) \, T^{(f)}_{ab} \, .
\ee
Note that, without loss of generality, we can assume $f(\phi,X) > 0$. This argument leads to the following results.

\begin{theorem} \label{Thm-1}
In viable Horndeski gravity, if we define the effective stress-energy tensor of the $\phi$-fluid as
\be
\label{Tf}
T^{(f)}_{ab} :=\frac{1}{f} \left(G_{ab} - \frac{1}{G_4} \, \mathcal{E}_{ab} \right) \, ,
\ee
with $f = f(\phi,X)$ an arbitrary strictly positive continuous function of $\phi$ and $X$, then the constitutive laws of the $\phi$-fluid can always be mapped onto those of Eckart's first-order thermodynamics. Furthermore, the corresponding thermal conductivity and temperature of gravity read
\begin{equation}
\label{eq:ktf}
{\cal K}^{(f)} {\cal T}^{(f)} = [f(\phi ,  X)]^{-1} \, \frac{\epsilon\sqrt{2X}\left(G_{4\phi}-XG_{3X}\right)}{G_{4}} \, ,
\end{equation}
whereas the evolution equation for the temperature of gravity for the $\phi$-fluid reads
\be
\label{eq:evolution-1}
  \begin{split}
    \frac{d\big[{\cal K}^{(f)} {\cal T}^{(f)}\big]}{d\tau} 
    &= \left[\epsilon \sqrt{2X} f_{\phi}- (\epsilon\sqrt{2X}\Box\phi-\Theta)f_{X} + \left(\epsilon\dfrac{\Box\phi}{\sqrt{2X}}-\Theta\right) f  \, \right] \frac{{\cal K}^{(f)} {\cal T}^{(f)}}{f}\\
    & \quad + \frac{1}{f} \, \nabla^c\phi\nabla_c\left(\dfrac{G_{4\phi}-XG_{3X}}{G_4}\right)
  \end{split}
\ee
with $\tau$ the proper time of an observer comoving with the $\phi$-fluid.
\end{theorem}

\begin{remark}
It is easy to see that for $f=1$ we recover the standard description of the first-order thermodynamics of scalar-tensor gravity.
\end{remark}
\begin{corollary} \label{Cor-1}
In viable Horndeski gravity, if we define the effective stress-energy tensor of the $\phi$-fluid as in \eqref{Tf} with $f = 1/G_4 \equiv G_{\rm eff}$, then
\begin{equation}
{\cal K}^{(G_{\rm eff})} {\cal T}^{(G_{\rm eff})} = \epsilon\sqrt{2X}\left(G_{4\phi}-XG_{3X}\right) \, ,
\end{equation}
and we also have that
\be
\label{eq:evolution-2}
  \begin{split}
    \frac{d\big[{\cal K}^{(G_{\rm eff})} {\cal T}^{(G_{\rm eff})}\big]}{d\tau} 
    &= \left[- \epsilon \sqrt{2X} \frac{G_{4\phi}}{G_{4}} + \left(\epsilon\dfrac{\Box\phi}{\sqrt{2X}}-\Theta\right) \, \right] {\cal K}^{(G_{\rm eff})} {\cal T}^{(G_{\rm eff})}\\
    & \quad + G_4 \, \nabla^c\phi\nabla_c\left(\dfrac{G_{4\phi}-XG_{3X}}{G_4}\right) \, ,
  \end{split}
\ee
with $\tau$ the proper time of an observer comoving with the $\phi$-fluid.
\end{corollary}

These results are particularly useful since they allow a simplification of the analysis of the approach to equilibrium equation for ``traditional'' scalar-tensor theories, {\em i.e.},
\be
S^{\rm (g)} _{\rm st} = \frac{1}{2} \int d^4x \sqrt{-g} \left[ \phi R 
-\frac{\omega(\phi )}{\phi} 
\, \nabla^c\phi \nabla_c\phi -V(\phi) \right] \,, \label{STaction}
\ee
which represents a well-investigated subclass of viable Horndeski gravity; specifically corresponding to 
$$
G_4 = \phi \, , \quad  G_2 = \frac{2 \ \omega (\phi)}{\phi} \, X - V (\phi)\, , \quad G_3 = 0 \, . 
$$
In fact, we shall show the following. 
\begin{theorem} \label{Thm-2}
In ``traditional'' scalar-tensor theories, for $f = 1/\phi \equiv G_{\rm eff}$ the evolution equation for the temperature of gravity reads
\be
\label{STApproachEq}
\frac{d\big[{\cal K}^{(G_{\rm eff})} {\cal T}^{(G_{\rm eff})}\big]}{d\tau} 
= - \Theta \, {\cal K}^{(G_{\rm eff})} {\cal T} ^{(G_{\rm eff})} + \Box\phi \, ,
\ee
with $\tau$ the proper time of an observer comoving with the $\phi$-fluid.
\end{theorem}

This result allows for a simplified proof of the fact that {\em electrovacuum} scalar-tensor theories satisfying $\omega = \mbox{const.}$, $V(\phi) = 0$, and $\Box \phi = 0$ will produce extreme deviations from General Relativity (diverging temperature of gravity) near spacetime singularities~\cite{Faraoni:2021lfc, Faraoni:2021jri}. Furthermore, Eq.~\eqref{STApproachEq} allows for a streamlined investigation of peculiar fixed points,  other than General Relativity, and their thermal stability~\cite{Giardino:2023qlu}.

\subsection{Structure of the work}

The manuscript is organized as follows. In Sec.~\ref{sec-2} we summarize the main results of the effective fluid approach and of the first-order thermodynamics of viable Horndeski gravity. In Sec.~\ref{sec-3} we provide the proofs of {\bf Theorem \ref{Thm-1}} and {\bf Corollary \ref{Cor-1}}. In Sec.~\ref{sec-4} we specialize the analysis presented in Sec.~\ref{sec-3} to ``traditional'' scalar-tensor theories and discuss the proof of {\bf Theorem \ref{Thm-2}}. Sec.~\ref{sec-5} discusses an alternative definition of the effective stress-energy tensor for viable Horndeski gravity, mixing the contributions of both matter and the scalar field, which is covariantly conserved and yielding a thermodynamic analogy with Eckart's theory only in {\em vacuo}. Last, in Sec.~\ref{sec-conc} we present some concluding remarks.

\subsection*{Acknowledgments}
M.M. is grateful for the support of Istituto Nazionale di Fisica Nucleare (INFN) {\em I.S. MOONLIGHT2}. The work of A.G. is supported in part by the Science and Technology Facilities Council (grants numbers ST/T006048/1 and ST/Y004418/1). A.~Mentrelli is partially supported by the MUR under the PRIN2022 PNRR project n.~P2022P5R22A. This work has been carried out in the framework of the activities of the Italian National Group of Mathematical Physics [Gruppo Nazionale per la Fisica Matematica (GNFM), Istituto Nazionale di Alta Matematica (INdAM)].

\newpage

\section{Preliminaries on viable Horndeski gravity}
\label{sec-2}

In this section, we shall briefly summarize the main results concerning the imperfect fluid representation of viable Horndeski gravity and the corresponding standard thermodynamic analogy with Eckart's theory. For details and derivations of the following results, we refer the reader to \cite{Giusti:2021sku,Miranda:2022wkz}.

Viable Horndeski gravity is a scalar-tensor theory defined by the action given in Eq.~\eqref{eq:viable}. This model provides the most general modified theory of gravity, built out of the metric tensor and a scalar field, leading to second-order field equations and the luminal propagation of gravitational waves.

Let us consider a scalar field $\phi$ with a timelike gradient. Taking advantage of the definition in Eq.~\eqref{4vel} one can easily compute the {\em kinematic quantities associated with the $\phi$-fluid}. Indeed, recalling the definitions of expansion scalar, shear tensor, and 4-acceleration, {\em i.e.} respectively,
\begin{eqnarray}
\Theta &\!\!:=\!\!& \nabla _a u^a \, ,\\
\sigma _{ab} &\!\!:=\!\!& \nabla_{(a}u_{b)} - \frac{\theta}{3} \, h_{ab} \, , \\
\dot u^a &\!\!:=\!\!& u^c \nabla_c u^a \, ,
\end{eqnarray}
and then it is easy to show that
\begin{eqnarray}
    \Theta &\!\!=\!\!&  
    \frac{\epsilon}{\sqrt{2 X}} \left( \Box \phi - \frac{\nabla X \cdot 
    \nabla \phi}{2 X} \right) \, , \label{theta}\\
    \sigma _{ab} &\!\!=\!\!& \frac{\epsilon}{\sqrt{2 X}} \left[ \nabla _a \nabla_b \phi - 
\frac{\nabla_{(a} X \nabla _{b)} \phi}{X} - \frac{\nabla X \cdot \nabla \phi}{4 X^2} \nabla _a \phi \nabla _b \phi - \frac{h_{ab}}{3} \left( \Box \phi - \frac{\nabla X \cdot \nabla \phi}{2X} \right) \right] \, , \\
2X\,\dot{u}_{a} &\!\!=\!\!& \,-\dot{X}\,u_{a} - \nabla_{a}X \, , \\
\Box\phi &\!\!=\!\!& \epsilon\left(\sqrt{2X} \, \Theta+\frac{\dot{X} }{\sqrt{2X }}\right) \label{box} \, ,
\end{eqnarray}
with $\nabla X \cdot \nabla \phi := g^{ab} \nabla_a X \nabla_b \phi$ and $\dot{X} := u^c \nabla_c X$. Note that these quantities are the same for all scalar-tensor models since their computation relies only on the definition of the four-velocity of the $\phi$-fluid, {\em i.e.}, Eq.~\eqref{4vel}.  

The imperfect fluid decomposition [Eq.~\eqref{eq:imperf}] of the effective energy-momentum tensor $T^{\rm (eff)}_{ab}$ for viable Horndeski gravity then yields
\begin{align}\label{tphiimp}
      T^{(\rm eff)}_{ab} =\, & \Bigg[ \frac{2XG_{2X}-G_{2} -2XG_{3\phi}}{2G_4}+\epsilon\frac{\sqrt{2X}  \left( G_{4\phi} -  X G_{3X} \right)}{  
    G_{4}}\,\Theta \Bigg]u_{a} u_{b}\nn\\
    &+ \Bigg[\frac{1}{2G_4} \left( G_2-2XG_{3\phi}+4XG_{4\phi\phi} \right)-\frac{\left(G_{4\phi}-XG_{3X}\right)}{G_4}\,\Box{\phi}+\epsilon\frac{\left(G_{4\phi}-3XG_{3X}\right)}{3G_4} \, \sqrt{2X}\,\Theta\, \Bigg] h_{ab}\nn\\
    &\nn\\
    &-\epsilon\frac{2\sqrt{2X} \left( G_{4\phi} - X G_{3X} \right)}{  G_{4}}\dot{u}_{(a} u_{b)}+\epsilon\frac{\sqrt{2X} \, G_{4\phi} }{ G_{4}} \, \sigma _{ab}\,.
\end{align}
which corresponds to identify the following effective fluid quantities:
\begin{align}
    \rho=\, & \frac{1}{2G_4}\left( 2XG_{2X}-G_{2} -2XG_{3\phi} \right)+\epsilon\frac{\sqrt{2X}}{G_4} \left( G_{4\phi}-XG_{3X} \right)\Theta\,,\label{rho}\\
    P=\,&\frac{1}{2G_4} \left( G_2-2XG_{3\phi}+4XG_{4\phi\phi} \right)-\frac{\left(G_{4\phi}-XG_{3X}\right)}{G_4}\,\Box{\phi}+\epsilon\frac{\left(G_{4\phi}-3XG_{3X}\right)}{3G_4} \, \sqrt{2X}\,\Theta\,,\label{P}\\
    q_{a}=\,&-\epsilon\frac{\sqrt{2X} \left( G_{4\phi} - X G_{3X} \right)}{  G_{4}}\dot{u}_{a}\,,\label{q}\\
    \pi_{ab}=\,&\epsilon\frac{\sqrt{2X} \, G_{4\phi} }{G_{4}} \, \sigma _{ab}\,.\label{pi}
\end{align}
Comparing these expressions with Eckart's constitutive equations [{\em i.e.}, Eq.~\eqref{EckartPvis}--\eqref{Eckartpi}] allows for the identification of the notions of the temperature of modified gravity and that of shear viscosity of modified gravity, {\em i.e.} respectively,
\be
{\cal K} {\cal T} := \frac{\epsilon\sqrt{2X}\left(G_{4\phi}-XG_{3X}\right)}{G_{4}} \quad \mbox{and} \quad \eta := - \epsilon\frac{\sqrt{2X} \, G_{4\phi} }{2 G_{4}} \, .
\ee
The identification of the bulk viscosity is not as straightforward; for details on the bulk viscous pressure and corresponding viscosity for viable Horndeski gravity, we refer the reader to \cite{Miranda:2022wkz} and \cite{Miranda:2024dhw}.

The evolution equation for the temperature then reads \cite{Miranda:2024dhw}
\be\label{eq:equilibriumH}
    \dfrac{{\rm d}({\cal KT})}{{\rm d} \tau}=\left(\epsilon\dfrac{\Box\phi}{\sqrt{2X}}-\Theta\right){\cal KT}+\nabla^c\phi\nabla_c\left(\dfrac{G_{4\phi}-XG_{3X}}{G_4}\right) \, .
\ee

\section{Alternative fluid representations: proof of Theorem \ref{Thm-1}}
\label{sec-3}

In this section, we shall provide detailed proofs of {\bf Theorem \ref{Thm-1}} and {\bf Corollary \ref{Cor-1}}. 

Again, let us consider viable Horndeski gravity with a scalar field $\phi$ such that its four-gradient is timelike and define the future-directed four-velocity $u^a$ associated with $\phi$ as in \eqref{4vel}. If we now consider the modified definition \eqref{eff-tensor-f} for the effective energy-momentum tensor of the $\phi$-fluid, it is easy to see that the effective Einstein equations for viable Horndeski gravity read as in Eq.~\eqref{eff-einst-f}. On the one hand, since the modification of the effective energy-momentum tensor of the $\phi$-fluid occurs only at the level of the effective Einstein equations, this procedure does not affect the kinematic quantities, which therefore remain identical to the one computed in Eq.~\eqref{theta}--\eqref{box}. On the other hand, because of Eq.~\eqref{eff-tensor-f} we have that the $\phi$-fluid still admits an imperfect fluid decomposition for $T^{(f)}_{ab}$, {\em i.e.},
\be 
T^{(f)} _{ab} = \rho^{(f)} u_{a}u_{b}+ P^{(f)} h_{ab}+2 q^{(f)} _{(a}u_{b)} +\pi^{(f)} _{ab} \, ,
\ee
with fluid quantities proportional to those in Eqs.~\eqref{rho}--\eqref{pi}, specifically,
\begin{align} 
\rho &= T ^{\rm (eff)} _{ab} u^a u^b = f \, T ^{(f)} _{ab} u^a u^b = f \, \rho^{(f)} \, , \label{eq:rhophi}\\ 
q_a &= - T ^{\rm (eff)} _{ab} \, u^c {h_a}^d = f \, q^{(f)}_a \, ,
  \label{eq:qphi}\\
 \Pi_{ab} &= P h_{ab} + \pi_{ab} = T ^{\rm (eff)} _{cd} \, {h_a}^c \, {h_b}^d = f \, \Pi^{(f)}_{ab} \, , 
\label{eq:Piphi}\\
    P &= \frac{1}{3}\, g^{ab}\Pi_{ab} =\frac{1}{3} \, h^{ab} T ^{\rm (eff)} = f \, P^{(f)} \, ,
\label{eq:Pphi}\\
    \pi_{ab} &= \Pi_{ab} - P h_{ab} = f \, \pi^{(f)}_{ab} \, \label{eq:piphi} \, ,
\end{align}
or in a more compact form $\{ \rho^{(f)} \, , \,\, P^{(f)} \, , \,\, q^{(f)}_a \, , \,\, \pi^{(f)} _{ab} \} = f^{-1} \, \{ \rho \, , \,\, P \, , \,\, q_a \, , \,\, \pi_{ab} \}$. Eqs.~\eqref{rho}--\eqref{pi} then imply that the constitutive relations of the $\phi$-fluid still map into those of Eckart's first-order thermodynamics up to a rescaling of the temperature and viscosities by a factor $1/f$. Let us consider the constitutive relation for the heat-flux density as an illustrative example. From Eqs.~\eqref{q} and~\eqref{eq:qphi} we can conclude that
\be 
q^{(f)}_a = -\epsilon\frac{\sqrt{2X} \left( G_{4\phi} - X G_{3X} \right)}{ f \,   G_{4}}\dot{u}_{a} \, ,
\ee
which upon comparing this expression with that of Eckart's theory [Eq.~\eqref{Eckartq}] implies a definition of ``temperature of modified gravity'' that reads
\be 
{\cal K}^{(f)} {\cal T}^{(f)} =  \frac{\epsilon\sqrt{2X}\left(G_{4\phi}-XG_{3X}\right)}{f \, G_{4}} \, ,
\ee
or, equivalently, ${\cal K}^{(f)} {\cal T}^{(f)} = {\cal K} {\cal T} / f$.

We can now investigate the evolution of this alternative definition of temperature as a function of the proper time of an observer comoving with the fluid. Specifically one has that
\be
  \begin{split}
   \frac{\d \left( {\cal K}^{(f)} {\cal T}^{(f)}\right)}{\d \tau} &= \,  u^c \nabla_c\left( {\cal K}^{(f)} {\cal T}^{(f)}\right)=u^c \nabla_c (f^{-1} \, {\cal K} {\cal T)} \\ 
   &= \frac{\K\T }{f^2} \left(\epsilon \sqrt{2X} f_{\phi}- \dot{X} \, f_{X} \right)+\frac{u^a \nabla_{a}(\K\T)}{f}\\ 
   &= \frac{\K\T }{f^2}\left[\epsilon \sqrt{2X}   f_{\phi}- (\epsilon\sqrt{2X}\Box\phi-\Theta)f_{X} \right]+\frac{1}{f} \frac{\d \left( {\cal K} {\cal T}\right)}{\d \tau} \, .
  \end{split}
\ee
Replacing $\frac{\d \left( {\cal K} {\cal T}\right)}{\d \tau}$ with the expression provided by Eq.~\eqref{eq:equilibriumH} yields
\be
  \begin{split}
\frac{\d \left( {\cal K}^{(f)} {\cal T}^{(f)}\right)}{\d \tau} &= \frac{\K\T }{f^2}\left[\epsilon \sqrt{2X}   f_{\phi}- (\epsilon\sqrt{2X}\Box\phi-\Theta)f_{X} \right] +\\
& \quad + \left(\epsilon\dfrac{\Box\phi}{\sqrt{2X}}-\Theta\right)\frac{{\cal K} {\cal T}}{f} + \frac{1}{f} \nabla^c\phi\nabla_c\left(\dfrac{G_{4\phi}-XG_{3X}}{G_4}\right) \, .
 \end{split}
\ee
If we now recall that ${\cal K}^{(f)} {\cal T}^{(f)} = {\cal K} {\cal T} / f$, we have that
\be
  \begin{split}
    \frac{d\big[{\cal K}^{(f)} {\cal T}^{(f)}\big]}{d\tau} 
    &= \left[\epsilon \sqrt{2X} f_{\phi}- (\epsilon\sqrt{2X}\Box\phi-\Theta)f_{X} + \left(\epsilon\dfrac{\Box\phi}{\sqrt{2X}}-\Theta\right) f  \, \right] \frac{{\cal K}^{(f)} {\cal T}^{(f)}}{f}\\
    & \quad + \frac{1}{f} \, \nabla^c\phi\nabla_c\left(\dfrac{G_{4\phi}-XG_{3X}}{G_4}\right) \, ,
  \end{split}
\ee
and this concludes the proof of {\bf Theorem \ref{Thm-1}}.

If we now choose $f = 1/ G_4$, {\em i.e.} we assume that gravity couples with the same strength to both matter and the effective $\phi$-fluid (somewhat implementing the weak equivalence principle at the level of the effective Einstein equations), from Eq.~\eqref{eq:ktf} we have that
\be 
{\cal K}^{(f)} {\cal T}^{(f)} \Big|_{f=1/ G_4} =  \frac{\epsilon\sqrt{2X}\left(G_{4\phi}-XG_{3X}\right)}{f \, G_{4}} \Bigg|_{f=1/ G_4} = \epsilon\sqrt{2X}\left(G_{4\phi}-XG_{3X}\right) \, ,
\ee 
and similarly
\be
  \begin{split}
    \frac{d\big[{\cal K}^{(f)} {\cal T}^{(f)}\big]}{d\tau} \Bigg|_{f=1/ G_4}  
    &= \left[- \epsilon \sqrt{2X} \frac{G_{4\phi}}{G_{4}} + \left(\epsilon\dfrac{\Box\phi}{\sqrt{2X}}-\Theta\right) \, \right] \left( {\cal K}^{(f)} {\cal T}^{(f)} \Big|_{f=1/ G_4} \right)\\
    & \quad + G_4 \, \nabla^c\phi\nabla_c\left(\dfrac{G_{4\phi}-XG_{3X}}{G_4}\right) \, ,
  \end{split}
\ee
thus concluding the proof of {\bf Corollary \ref{Cor-1}}.

\section{``Traditional'' scalar-tensor theories: proof of Theorem \ref{Thm-2}}
\label{sec-4}

To prove the statement in {\bf Theorem \ref{Thm-2}} we have to specialize our analysis of viable Horndeski gravity to its subclass that coincides with ``traditional'' scalar-tensor theories, {\em i.e.}, the subclass such that 
$$
G_4 = \phi \, , \quad  G_2 = \frac{2 \ \omega (\phi)}{\phi} \, X - V (\phi)\, , \quad G_3 = 0 \, ,
$$
with $\phi > 0$ over the spacetime manifold.

Taking advantage of {\bf Corollary \ref{Cor-1}} we can easily infer that
\be 
{\cal K}^{(f)} {\cal T}^{(f)} \Big|_{f=\frac{1}{\phi} \, , \, {\rm st}} = \frac{\epsilon\sqrt{2X}\left(G_{4\phi}-XG_{3X}\right)}{f \, G_{4}}\Bigg|_{f=\frac{1}{\phi} \, , \, {\rm st}} = \epsilon\sqrt{2X} \, .
\ee
Furthermore, from Eq.~\eqref{eq:evolution-1} specialized to ``traditional'' scalar-tensor theories we have that
\be
  \begin{split}
    \frac{d\big[{\cal K}^{(f)} {\cal T}^{(f)}\big]}{d\tau} \Bigg|_{f=\frac{1}{\phi} \, , \, {\rm st}}  
    &= \left[- \frac{\epsilon \sqrt{2X}}{\phi}  + \left(\epsilon\dfrac{\Box\phi}{\sqrt{2X}}-\Theta\right) \, \right] \left( {\cal K}^{(f)} {\cal T}^{(f)} \Big|_{f=\frac{1}{\phi} \, , \, {\rm st}} \right) + \phi \, \nabla^c\phi\nabla_c\left(\dfrac{1}{\phi}\right) \\
    &= \left[- \frac{\epsilon \sqrt{2X}}{\phi}  + \left(\epsilon\dfrac{\Box\phi}{\sqrt{2X}}-\Theta\right) \, \right] \, \epsilon\sqrt{2X} + \frac{2X}{\phi} \\
    &= \left(\epsilon\dfrac{\Box\phi}{\sqrt{2X}}-\Theta\right) \epsilon\sqrt{2X} 
    = \Box\phi - \Theta \, \epsilon\sqrt{2X} \\
    &= \Box\phi - \Theta \, \left( {\cal K}^{(f)} {\cal T}^{(f)} \Big|_{f=\frac{1}{\phi} \, , \, {\rm st}} \right) \, ,
  \end{split}
\ee
which concludes the proof of {\bf Theorem \ref{Thm-2}}.

%
\section{An effective fluid with covariantly conserved stress-energy tensor}
\label{sec-5}
%

As pointed out in Sec.~\ref{sec-1.1}, the effective stress-energy tensor~\eqref{tphiimp} is not covariantly conserved due to the nonminimal coupling of the scalar field to Einstein's gravity through $G_4 (\phi)$ [Eq.~\eqref{nonconserved}]. Nonetheless, one could try to look for alternative definitions for the effective stress-energy tensor that are covariantly conserved and see their effects on the thermodynamic analogy.

Turning our attention to viable Horndeski gravity, the simplest way of constructing a conserved rank-two tensor consists in considering linear combinations of conserved quantities. For instance, we know that $\mathcal{E}_{ab}$, as defined in Eq.~\eqref{fieldeq}, and $G_{ab}$ are divergence-free tensors (on-shell), thus a simple proposal for an alternative covariantly-conserved stress-energy tensor reads
\begin{equation}\label{conserved}
    \hat{T}_{ab}:=G_{ab}-\mathcal{E}_{ab}\,,
\end{equation}
leading to the following form of the effective Einstein equation
\begin{equation}\label{conserved_effeqs}
    G_{ab}=\hat{T}_{ab}+T_{ab}^{\rm(m)}\,.
\end{equation}
A nice feature of this definition is that both matter and the ``effective fluid'' are equally coupled to Einstein's gravity. However, this is done at the price of mixing matter and scalar field contributions, and this leads to complications in the thermodynamic analogy. Indeed, the conserved effective stress-energy tensor $\hat{T}_{ab}$ can be written in terms of the \textit{minimal} one ${T}^{\rm (eff)}_{ab}$ [Eq.~\eqref{tphiimp}] as
\begin{equation}
    \hat{T}_{ab}={T}^{(\rm eff)}_{ab}+\left(\frac{1-G_{4}}{G_{4}}\right)T^{\rm(m)}_{ab}\,.
\end{equation}
Then, assuming that the scalar field has a timelike gradient and defining the effective fluid 4-velocity as in Eq.~\eqref{4vel} we can perform an imperfect fluid decomposition of $\hat{T}_{ab}$ which yields 
\begin{align}
    \hat{\rho}=&\,\rho+\left(\frac{1-G_{4}}{G_{4}}\right)T^{\rm(m)}_{ab}u^{a}u^{b}\,,\\
    &\nn\\
    \hat{P}=&\,P+\frac{1}{3}\left(\frac{1-G_{4}}{G_{4}}\right)T^{\rm(m)}_{ab}h^{ab}\,,\\
    &\nn\\
    \hat{q}_a=&\,q_a-\left(\frac{1-G_{4}}{G_{4}}\right)T^{\rm(m)}_{cd}u^{c}h^{d}{}_{a}\,,\\
    &\nn\\
    \hat{\pi}_{ab}=&\,\pi_{ab}+\left(\frac{1-G_{4}}{G_{4}}\right)T^{\rm(m)}_{ab}\left(h^{ac} h^{bd} -\frac{1}{3}h^{cd}h^{ab} \right)\, ,
\end{align}
where $\rho$, $P$, $q_a$, and $\pi_{ab}$ as in Eq.~\eqref{rho}--\eqref{pi}.

Because of the presence of matter in the definition of the effective stress-energy tensor the fluid quantities read as the ones resulting from a mixture of tilted fluids~\cite{King:1972td, Miranda:2022uyk}. Only in the case of a minimally coupled scalar field, {\it i.e.} $G_{4}=1$ corresponding to the so-called kinetic gravity braiding, the proposed definition is equivalent to that of ${T}^{\rm (eff)}_{ab}$ [Eq.~\eqref{tphiimp}].

It is now crucial to note that, in light of the above observations, the thermodynamic analogy with Eckart's first-order thermodynamics holds only for $T^{(\rm eff)}_{ab}$ and $T^{(f)}_{ab}$, for which the associated quantities do not explicitly contain matter contributions, and not for the alternative definition in Eq.~\eqref{conserved}. Hence, such a definition would have limited use within the thermodynamics of scalar-tensor gravity.
%
%
\section{Discussion of the results}
\label{sec-conc}
%

We examined the foundations of the formalism underlying the thermodynamics of scalar-tensor gravity. The exploration of the properties of the imperfect fluid decomposition for modified theories of gravity led us to a class of alternative formulations of the thermodynamic analogy connecting scalar-tensor theories with Eckart's first-order thermodynamics. In particular, we made clear the need for the identification of an effective stress-energy tensor such that standard matter does not mix with the additional (non-minimally coupled) scalar field contribution. The only additional free parameter, that does not spoil the thermodynamic analogy, turns out to be the ``coupling'' $f(\phi, X)$ of this effective stress-energy tensor for the $\phi$-fluid to Einstein's gravity.

More in detail, in Sec.~\ref{sec-3} we have shown that if we reformulate the thermodynamic formalism for an effective stress-energy tensor for the $\phi$-fluid defined as in Eq.~\eqref{eff-tensor-f}, then this identification preserves the form of the constitutive relations for the $\phi$-fluid. This allows for an identification of the fluid quantities analogous to the standard formalism, up to a rescaling of a factor $[f(\phi, X)]^{-1}$, and to a slightly altered evolution equation for the temperature of gravity. Furthermore, when restricted to the ``traditional'' class of scalar-tensor theories the identification $f(\phi)=1/\phi$ allows for a streamlined implementation of the original thermodynamic analogy, of the investigation of the evolution of the temperature of gravity, and it has also the additional perk that the $\phi$-fluid couples to Einstein's gravity with the same strength as the matter fields.

These results not only reinforce the connection between alternative theories of gravity and non-equilibrium thermodynamics, but they also open new avenues for exploring peculiar thermal states and the thermal stability of these gravitational theories. Future work will focus on further investigating the implications of this thermodynamic analogy, particularly in the context of cosmological and astrophysical scenarios, where the effects of scalar fields could play a role.

\end{document}